\documentclass[fleqn,usenatbib]{mnras}

\usepackage{newtxtext,newtxmath}

\usepackage[T1]{fontenc}

\DeclareRobustCommand{\VAN}[3]{#2}
\let\VANthebibliography\thebibliography
\def\thebibliography{\DeclareRobustCommand{\VAN}[3]{##3}\VANthebibliography}

\usepackage{graphicx}	
\usepackage{amsmath}	
\usepackage{pdflscape}

\title[Clusters, Clouds, Correlations: M33 and M31]{Clusters, Clouds, and Correlations: \\Relating Young Clusters to Giant Molecular Clouds in M33 and M31}

\author[J. Peltonen et al.]{Joshua Peltonen,$^{1}$\thanks{E-mail: peltonen@ualberta.ca}
Erik Rosolowsky,$^{1}$
L. Clifton Johnson,$^{2}$
Anil C. Seth,$^{3}$
Julianne Dalcanton,$^{4,5}$
\newauthor Eric F.\ Bell$^{6}$
Jonathan Braine,$^{7}$
Eric W. Koch,$^{8}$
Margaret Lazzarini,$^{9}$
Adam K. Leroy,$^{10,11}$
Evan D. Skillman,$^{12}$
\newauthor Adam Smercina,$^{4}$
Tobin Wainer,$^{4,3}$
Benjamin F. Williams$^{4}$\\
$^{1}$Department of Physics, University of Alberta, Edmonton, AB T6G 2E1, Canada\\
$^{2}$Center for Interdisciplinary Exploration and Research in Astrophysics (CIERA) and Department of Physics and Astronomy, Northwestern University,\\
1800 Sherman Ave., Evanston, IL 60201, USA\\
$^{3}$Department of Physics and Astronomy, University of Utah, Salt Lake City, UT 84112, USA\\
$^{4}$Department of Astronomy, University of Washington, Box 351580, U.W., Seattle, WA 98195-1580, USA\\
$^{5}$Center for Computational Astrophysics, Flatiron Institute, 162 Fifth Avenue, New York, NY 10010, USA\\
$^{6}$Department of Astronomy, University of Michigan, 1085 S. University Avenue, Ann Arbor, MI, 48105, USA\\
$^{7}$Laboratoire d’Astrophysique de Bordeaux, Univ. Bordeaux, CNRS, B18N, Allée Geoffroy Saint-Hilaire, 33615 Pessac, France\\
$^{8}$Center for Astrophysics $\mid$ Harvard \& Smithsonian, 60 Garden St., 02138 Cambridge, MA, USA\\
$^{9}$Division of Physics, Mathematics and Astronomy, California Institute of Technology, 1200 E California Boulevard, Pasadena, CA 91125, USA\\
$^{10}$Department of Astronomy, The Ohio State University, 140 West 18th Avenue, Columbus, OH 43210, USA\\
$^{11}$Center for Cosmology and Astroparticle Physics, 191 West Woodruff Avenue, Columbus, OH 43210, USA\\
$^{12}$Minnesota Institute for Astrophysics, 116 Church Street SE, Minneapolis, MN, 55455, USA
}

\date{Accepted XXX. Received YYY; in original form ZZZ}

\pubyear{2023}

\begin{document}
\label{firstpage}
\pagerange{\pageref{firstpage}--\pageref{lastpage}}
\maketitle

\begin{abstract}
We use young clusters and giant molecular clouds (GMCs) in the galaxies M33 and M31 to constrain temporal and spatial scales in the star formation process. In M33, we compare the PHATTER catalogue of 1214 clusters with ages measured via colour-magnitude diagram (CMD) fitting to 444 GMCs identified from a new 35~pc resolution ALMA $^{12}$CO(2-1) survey. In M31, we compare the PHAT catalogue of 1249 clusters to 251 GMCs measured from a CARMA $^{12}$CO(1-0) survey with 20~pc resolution. Through two-point correlation analysis, we find that young clusters have a high probability of being near other young clusters, but correlation between GMCs is suppressed by the cloud identification algorithm. By comparing the positions, we find that younger clusters are closer to GMCs than older clusters. Through cross-correlation analysis of the M33 cluster data, we find that clusters are statistically associated when they are $\leq$10~Myr old. Utilizing the high precision ages of the clusters, we find that clusters older than $\approx 18$~Myr are uncorrelated with the molecular ISM. Using the spatial coincidence of the youngest clusters and GMCs in M33, we estimate that clusters spend $\approx$4-6~Myr inside their parent GMC. Through similar analysis, we find that the GMCs in M33 have a total lifetime of $\approx 11$-15~Myr. We also develop a drift model and show that the above correlations can be explained if the clusters in M33 have a 5-10~km~s$^{-1}$ velocity dispersion relative to the molecular ISM.
\end{abstract}




\section{Introduction}

The essential heavy elements around us are evidence of the cycling process between gas and stars in the Universe. Molecular gas in the interstellar medium (ISM) forms high-mass stars and clusters that, in their evolution, will disrupt the surrounding gas. This star formation and resulting disruption rely on the interplay of gravity, turbulence, stellar feedback, magnetic fields, chemistry, and thermal regulation. The processes that play the most significant role in regulating star formation remain areas of active research \citep{2007MckeeandOs,2022Chevance}. 

Constraining the timescales associated with the various phases of the star formation process can help constrain which physical processes are at play. This has motivated a number of studies that attempt to infer the evolutionary timescales of the molecular ISM, in particular, lifetimes of giant molecular clouds (GMCs) or the timescale over which feedback destroys the GMCs. GMCs are massive collections of molecular gas that are the primary sites of star formation. Early attempts to measure the lifetimes of GMCs led to a range of results with some estimates converging to long lifetimes \citep[$10^{7.5}-10^{8}$~yr;][]{1977Bash,1979Scoville} and others to shorter lifetimes \citep[$<10^{7.5}$~yr;][]{1980Blitz,1993Blitz}. Lifetimes that are $10^{7.5}$~yr or longer require GMCs to survive much of a galactic rotation period with forces that prevent gravitational collapse. Some recent analyses of gas distribution still point to long GMC lifetimes \citep{2012Koda,2016Koda,2020Koda}. However, there are also many approaches that point to short lifetimes. \citet{Kawamura2009} used the spatial coincidence of GMCs and young clusters to determine that GMCs live for 20-30~Myr in the Large Magellanic Cloud (LMC). Numerical simulations have also suggested a short molecular cloud lifetime, for example, 4-25~Myr in \citet{dobbs13} and 13-20~Myr in \citet{2021Jeffreson}. Using the ``tuning fork'' measurements of the decorrelation between star formation tracers and molecular gas developed in \citet[][see also \citealt{2010Onodera}]{2010Schruba},  \cite{2014Kruijssen} and \cite{2018Kruijssen} developed their ``uncertainty principle'' formalism to measure a short \citep[10~Myr in NGC 300][]{2019Kruijssen} lifetime for molecular clouds. This methodology has now been replicated in simulations \citep{2021Semenov} and applied to broader samples of galaxies \citep{2022ChevanceTuning,2021Kim,2022kim}. \cite{2022ChevanceTuning} argued for short lifetimes and contended that GMCs are dispersed quickly after the onset of star formation. They suggest that pre-supernova feedback mechanisms play a key role in disrupting clouds. 

While the details of some of the measurements can be quite sophisticated, these measurements are usually framed around our na\"ive model for star formation in molecular clouds. The model starts with overdense turbulent complexes of molecular gas that can be divided into individual GMCs. However, the boundaries between GMCs and the outside ISM are not always clear \citep{2022Chevance}. These GMCs then begin forming clusters that inherit the clustered structure of their progenitor clouds \citep{2018Grasha, 2019Grasha, 2022Turner}. After the cluster spends some time inside their progenitor clouds they will disrupt the cloud through a combination of supernovae and stellar feedback \citep[the significance of each effect is still debated;][]{2018Kim,2022ChevanceTuning}. We define the lifetime of a GMC as the time between when the cloud can first be detected to when it can no longer be detected because of this disruption. The disruption of the dense gas occurs before most of the gas can be converted into stars, leading to very long depletion times and inefficient star formation \citep{2022Chevance}. After the clusters have dispersed the gas, the correlated structure of the clusters will be erased by random drift velocities inherited from the turbulent motions of the gas.

Many of the approaches to characterize the evolutionary timescale of the molecular ISM rely on tracers of star formation (H$\alpha$, UV, and mid-infrared) that do not directly measure the ages of a stellar population. Instead, they trace the integrated radiation from a stellar population in a given waveband; for example, H$\alpha$ traces the ionizing photon radiation that comes from short-lived O-stars. Using an assumed model of the initial mass function, these measurements can then be translated back into star formation rates and characterize the timescale for evolution of molecular clouds, providing a narrow window into the star formation history \citep{kennicutt2012}. These tracers are typically restricted to only a portion of the star formation process. H$\alpha$, for example, requires high-mass stars and is thus restricted to star formation events that host massive star formation.

Alternatively, the ages of simple stellar populations provide a robust method to establish evolutionary timescales and provide a long view into the star formation history of a galaxy. Stellar clusters represent close approximations of simple stellar populations, which \citet{Kawamura2009} leveraged for their measurement of cloud evolution. Such studies cannot be carried out in the Milky Way for large samples because of line-of-sight blending and extinction. Therefore, comparing clusters and GMCs has been limited only to the nearest galaxies (e.g., the LMC) or limited to unresolved cluster candidates in galaxies like M33 \citep{2017CorbM33}. However, with recent \textit{Hubble Space Telescope} (HST) surveys and interferometer observations, resolved clusters can be compared to GMCs in nearby galaxies like is done in \citet{2018Grasha,2019Grasha} and \citet{2022Turner}.   

In this work, we take advantage of new, high-quality data to measure the relationship between cluster and molecular cloud populations in the two largest star-forming galaxies in the Local Group: M31 and M33. Clusters were identified in these galaxies using the wide-area HST surveys, PHAT \citep{2012ApJS..200...18D} in M31 and PHATTER \citep{2021PHATTERI} in M33. In addition, high-resolution surveys of $^{12}$CO, obtained by the Combined Array for Research in Millimeter-wave Astronomy (CARMA) in M31 \citep{calduprimo16,leroy16,schruba19} and the Atacama Large Millimeter/submillimeter Array (ALMA) in M33 (Koch et al. in prep), allow for the analysis of GMCs at sub-cloud scales. M31 is about three times the diameter \citep{1991Dbook} and 23$^\circ$ more inclined \citep{2018KochM33Inc,2019ApJ...872...24V} than M33. In M33, the surveys thoroughly cover the central part of the galaxy and a majority of the star-forming spiral arms \citep{2021PHATTERI}. Due to its inclination and large area on the sky, the HST surveys in M31 are more limited and only cover a quadrant of the galaxy. Despite the limitations in M31, these high-resolution surveys and the relative proximity of M31 and M33 allow for colour-magnitude diagram (CMD) fitting of the clusters, which yields robust age and mass estimates. These accurate ages and the relatively deep mass completeness limits allow for an unprecedented study of molecular gas and star formation. 

We focus on the correlation structure between the GMCs identified in the $^{12}$CO surveys and the clusters identified in the HST surveys. Previous studies have shown that very young clusters are typically near GMCs \citep{Kawamura2009,2014Whitmore,2017CorbM33,2018Grasha,2019Grasha}. It has also been shown that the two-point correlation is stronger for younger clusters than older clusters \citep{2015Grasha,2017Grasha,2021Menon}. Finally, the cross-correlation function has been used to show that young clusters are more correlated with GMCs than old clusters \citep{2022Turner}. However, \citet{2021Li} have noted that applying correlation functions to non-homogeneous populations may produce spurious correlation signal. While the more sophisticated ways of interpreting correlation, like spatial point processes, have their benefits, they are difficult to interpret and implement. Therefore, we have decided to address some of the concerns with correlation functions by building in the overall effects of galactic structure by using random cluster distributions that contain the same non-homogeneous structure.

The details of the M33 surveys and the more limited M31 surveys are presented in Section \ref{obs}. Because of the limitations of the M31 data, Section \ref{m33analy} focuses on the main results found in M33. These results include an analysis of the correlation between clusters and GMCs, a comparison of cluster and GMC properties, and estimations of GMC timescales. Section \ref{m31analy} is a partial parallel analysis on M31. In Section \ref{Discussion}, we develop a simple model to estimate the drift velocity of the clusters and discuss the effects of completeness. Finally, a summary of the work is presented in Section \ref{Conclusion}.

\section{Observations}\label{obs}

\begin{table}
 \caption{Adopted Parameters for M33 and M31.}
 \label{tab:galaxydetails}
 \begin{tabular}{lcc}
  \hline
   & M33 & M31 \\
  \hline
  Distance (kpc)& 859$^{a}$  & 776$^{b}$  \\
 
  Orientation & $i$=55$^\circ$$^{c}$, PA=201$^\circ$$^{c}$ & $i$=78$^\circ$$^{d}$, PA=38$^\circ$$^{d}$  \\

  Central position$^{e}$& RA=23.46204$^\circ$& RA=10.68479$^\circ$ \\
  & DEC=30.66022$^\circ$ & DEC=41.26907$^\circ$\\
  SFR$^{e}$ (M$_\odot$/yr) & 0.32& 0.39\\
  Stellar Mass$^{e}$ (M$_\odot$) & 2.63$\times10^{9}$ & 5.37$\times10^{10}$\\
  \hline
 \end{tabular}
 \\
 $^{a}$\cite{2017distanceM33};
 $^{b}$\cite{2022M31Dis}; $^{c}$\cite{2018KochM33Inc}; $^{d}$\cite{2019ApJ...872...24V}; $^{e}$\citet{2019Leroy}
\end{table}
\subsection{M33}
\begin{figure}
	\includegraphics[width=\columnwidth]{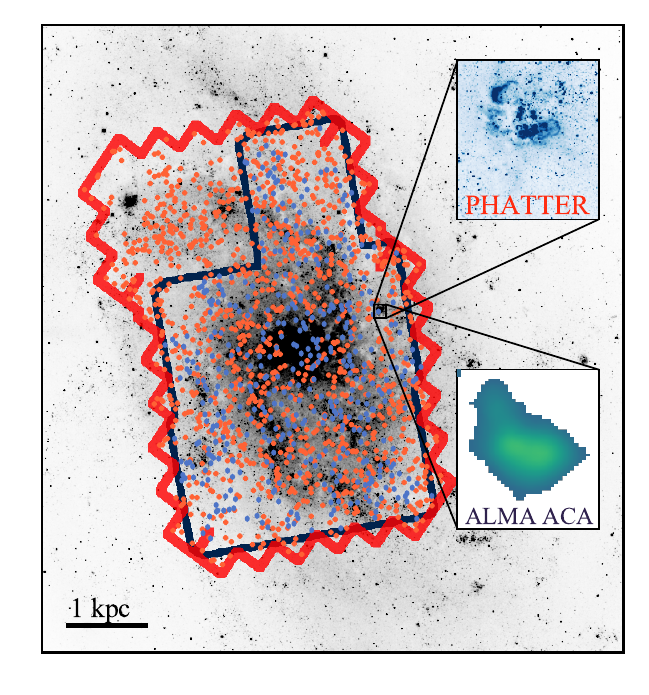}
    \caption{The data used to study the clusters and GMCs in M33. The background is a \textit{B} band image from the 4~m Mayall telescope \citep{2006Bband}. The red outline shows the footprint of the PHATTER survey, while the dark blue outline shows the footprint of the ALMA ACA survey. The top zoomed-in frame shows the resolution of the PHATTER data. The bottom zoomed-in frame shows a GMC in the same area identified from the ALMA ACA data. The orange circles show the locations of the clusters identified from the PHATTER data. The GMCs identified from the ALMA ACA survey are shown as blue diamonds.}
    \label{fig:m33dat}
\end{figure}

We trace GMCs in M33 using a new CO survey carried out using the ALMA Atacama Compact Array (ACA). The ALMA ACA survey covers a section of M33 shown in Figure \ref{fig:m33dat} as a dark blue outline and is described in more detail in Koch et al. (in prep). The ACA survey is centred on the $^{12}$CO J=2-1 transition at 230.538~GHz with a bandwidth of $154~\mathrm{MHz} \approx 200~\mathrm{km~s^{-1}}$. The synthesized beam size of $\approx8\farcs5$ corresponds to a physical size of $\approx35$~pc at the distance of M33 from Table \ref{tab:galaxydetails}. The data have a noise level of 45 mK in a 0.7~km~s$^{-1}$ channel. Koch et al. (in prep) applied the Spectral Clustering for Molecular Emission Segmentation (SCIMES) algorithm \citep{2015SCIMES} to the ACA data to obtain a catalogue of 444 GMCs. These GMCs are shown in Figure \ref{fig:m33dat} as blue diamonds. An integrated intensity map for one of these GMCs is shown in Figure \ref{fig:m33dat} in the bottom zoomed-in frame.

The PHATTER survey \citep{2021PHATTERI} is composed of observations in six HST filters and covers a region of M33 shown in Figure \ref{fig:m33dat} as a red outline. The top zoomed-in panel of Figure \ref{fig:m33dat} shows a portion of the PHATTER survey (in the F475W filter for HST), illustrating the resolution is sufficient to identify individual stars and clusters. Using a crowd-sourced visual search, \citet{2022PHATTERIV} identified 1214 clusters from the PHATTER survey that are believed to be long-lived. These clusters are marked in Figure \ref{fig:m33dat} as orange circles. The ages and masses of the clusters were identified using CMD fitting as described in \cite{2022PHATTERIII}. This catalogue of optically identified clusters is distinct from the catalogue of 630 young star cluster candidates (YSCCs) in \citet{2017CorbM33}  found using \textit{Spitzer Space Telescope} 24~$\mu$m images \citep{2007Verley}. These infrared YSCCs, originally identified by \cite{2011Sharma}, are thought to be young ($\leq$10~Myr) embedded objects. However, \cite{2022PHATTERIV} found that at least 30\% of the YSCCs cannot be embedded young clusters because of the lack of visual extinction. Because of the potential contamination of non-cluster objects, we will only use this YSCC catalogue from \citet{2017CorbM33} for comparison in Section \ref{2pnt} and \ref{cross}. 

\subsection{M31}
\begin{figure}
	\includegraphics[width=\columnwidth]{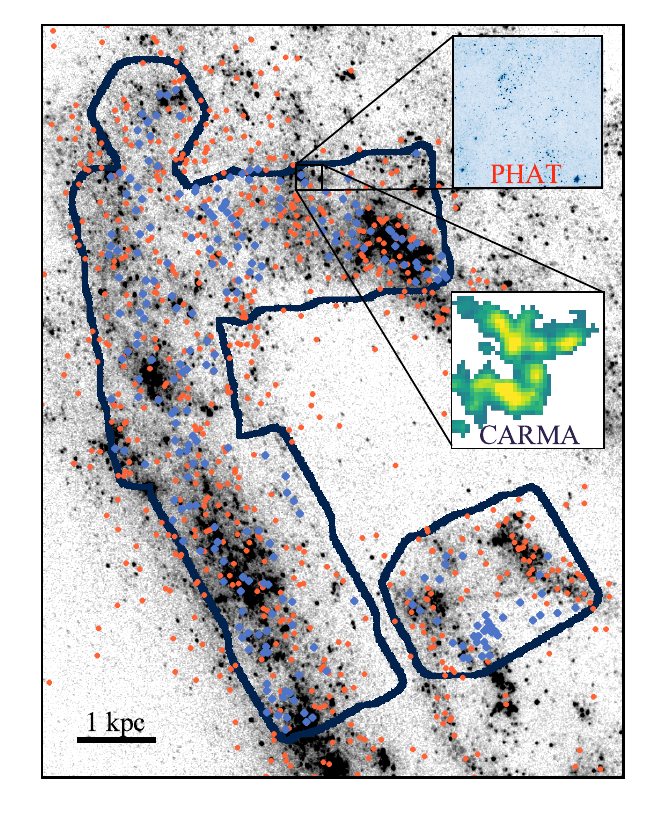}
    \caption{The data used to study the clusters and GMCs in M31. The background is a far-ultraviolet image from the \textit{Galaxy Evolution Explorer} \citep{2007Galex}. The dark blue outline shows the footprint of the CARMA survey. The top zoomed-in frame shows the resolution of the PHAT data. The bottom zoomed-in frame shows a GMC in the same area identified from the CARMA data. The orange circles show the locations of the clusters identified from the PHAT data. The GMCs identified from the CARMA survey are shown as blue diamonds.}
    \label{fig:m31dat}
\end{figure}

A portion of M31 was surveyed using CARMA, PI A. Schruba, which includes short spacing data from Nieten et al. (2006).  These data have appeared in \citet{calduprimo16}, \citet{leroy16}, and \citet{schruba19} with description and images of the data in those papers. Figure \ref{fig:m31dat} shows the area surveyed in the CO survey. With a synthesized beam size of $\approx5\farcs5$ and at the distance of M31 from Table \ref{tab:galaxydetails}, the physical resolution is $\approx20$~pc. The data have a noise level of 190 mK in a 2.5 km~s$^{-1}$ channel. We do not match the CARMA resolution to the ALMA ACA resolution because we want to utilize the highest resolution possible, and most of our analysis will be done at scales larger than the resolution.

We then applied the SCIMES algorithm \citep{2015SCIMES} to the CARMA data, which yielded a catalogue of 251 GMCs. We use the same default algorithm parameters as were used in Koch et al. (in prep) for signal identification and cloud decomposition. These GMCs are shown in Figure \ref{fig:m31dat} as blue diamonds. Figure \ref{fig:m31dat} also shows an integrated intensity map of one of the clouds in the bottom zoomed-in frame. 

We use the stellar cluster catalogue generated from the PHAT survey \citep{2012ApJS..200...18D}, the predecessor to PHATTER. PHAT covers a quadrant of M31's star-forming disk in six HST bands. Figure \ref{fig:m31dat} shows an area that is smaller than the PHAT survey, with a small portion shown in the top zoomed-in panel. Using a crowd-sourced visual search, \cite{2015ApJ...802..127J} identified 2753 clusters from the PHAT survey. Age and mass estimates are derived from CMD fitting, and a subsample of 1249 clusters with ages between 10-300~Myr were reported in \citet{2016ApJ...827...33J}.  These clusters are shown in Figure \ref{fig:m31dat} as orange circles. 

\subsection{Completeness}\label{sec:completeness}
Throughout the analysis, we will use all of the clusters identified in \citet{2022PHATTERIV} and \citet{2016ApJ...827...33J} to use the greatest number of sources. However, our results will depend on the completeness properties of the cluster and GMC catalogues. For the clusters in both PHATTER and PHAT, the visual search for clusters included injected synthetic clusters of known mass and age \citep{2015ApJ...802..127J,2022PHATTERIV}. These synthetic clusters allow for thorough completeness analysis. In M31, the cluster catalogue at 100~Myr is 50\% complete down to $M_\mathrm{c,50}\approx$500~M$_\odot$ \citep{2015ApJ...802..127J}. The completeness is better for younger clusters since they are more likely to have bright young stars. However, the oldest clusters in the catalogue (100-333~Myr) are 50\% complete down to $M_\mathrm{c,50}\approx$1000~M$_\odot$ \citep{2015ApJ...802..127J}. The completeness in M33's cluster catalogue is worse than in M31 with a 50\% completeness limit of $\approx$1000~M$_\odot$ at an age of 100 Myr \citep{2022PHATTERIV}. This difference in completeness in M31 and M33 comes from the higher degree of crowding in M33, which makes identifying clusters more difficult \citep{2022PHATTERIV}. Therefore, more crowded regions of each galaxy will also lead to lower completeness in these regions. Another property that has an impact on cluster completeness is extinction. The very youngest (<3~Myr) deeply embedded clusters will be difficult to identify due to optical extinction \citep{2022PHATTERIV}. The very young embedded stars should be visible with recent observations from the \textit{James Webb Space Telescope} (JWST), which can see with eyes unclouded by the optical extinction of GMCs (Peltonen et al., in prep). 

For M33 and M31, we measure the 50\% completeness limit for molecular cloud identification as $M_\mathrm{GMC,50}=1.3\times 10^{4}~M_\odot$ and $M_\mathrm{GMC,50}=3.0\times 10^4~M_\odot$ respectively, based on the artificial cloud recovery test method presented in \citet{rosolowsky21}. This approach inserts GMCs of known brightness and properties into the data and tests whether they are recovered in the cloud identification algorithm. This value is approximately 40$\times$ the $1\sigma$ noise level in a single synthesized beam. Our estimate assumes a Galactic CO-to-H$_2$ conversion factor \citep{bolatto13} and a CO(2-1)/CO(1-0) line ratio of $R=0.7$ \citep{leroy22}. 

Another issue to consider is not the absolute mass limit of the cloud and cluster catalogs independently, but possible mismatches between these two limits. In other words, could small clusters be undetectable when hosted by the lowest mass of the GMC sample, or conversely, could the host GMC of a low mass stellar cluster be undetectable in the CO observations? The mass of the GMC completeness limit is 15 to 30 times larger than the corresponding cluster mass completeness limit (typically $10^3~M_\odot$ as above), suggesting these two catalog limits are well-matched as long as the efficiency of a GMC forming a cluster (by mass) is $\sim$0.03-0.06.  If the true efficiency is lower (i.e., a given GMC can only host a smaller maximum-mass cluster), then some catalogued GMCs may host clusters that couldn't actually be detected, and the cluster catalog would be incomplete with respect to the GMC catalog.  If the efficiency is higher, then the converse is true and some low mass stellar clusters may actually live in GMCs that are too low mass to have been detected, and the GMC catalog would be incomplete with respect to the stellar cluster catalog. 

We can compare the ``matched'' catalog of efficiency of 0.03-0.06 to current constraints on the efficiency of stellar clusters forming in GMCs.  Estimates suggest that, on the scale of individual clouds, up to 30\% of the cloud mass may be converted into stars \citep{krumholz19}. This is a much higher efficiency, suggesting that the lowest mass stellar clusters in our samples may in fact be hosted by undetected GMCs, but that every cataloged GMC that hosts a stellar cluster should have the cluster detected.  In other words, the cluster catalog is complete with respect to the GMC catalog, but the GMCs are not complete with respect to the cluster catalog. We can also look at global, ensemble estimates of the efficiency, by multiplying the efficiency of turning molecular gas into stars on large (kpc) scales ($f_\star$) by the fraction of the overall star formation that produces stellar clusters ($f_\mathrm{clust}$). Current estimates of these quantities are $f_\star\sim 0.03$ \citep[][]{2018Utomo,2022Chevance,2022kim} and $f_\mathrm{clust}\sim 0.1$ \citep[][]{krumholz19}, giving an overall efficiency of $f_\star f_\mathrm{clust}\sim 0.003$.  This is much lower than our ideal "matched" catalog efficiency, which would imply that some GMCs in our catalog may actually host undetectable stellar clusters, but that every stellar cluster should have its host GMC detected. In this case, the strength of our cross-correlation signals would be lower limits since the presence of more clusters correctly matched near progenitor cloud structures should enhance the cross-correlation amplitude. 

We can also assess whether there is additional low mass incompleteness in the stellar cluster catalog by constraining the population of embedded clusters using the 24~$\mu$m-derived YSCC catalog from \citet{2017CorbM33} to assess whether there are infrared sources without associated CO emission, which would imply our survey misses clouds hosting embedded cluster formation.  We find that $188/244 = 77\%$ of sources in the survey area overlap with CO clouds.  Since up to 30\% of these infrared sources could be interlopers \citep{2022PHATTERIV}, the degree of spatial coincidence suggests our cloud catalogue is sufficiently deep to include most of the cluster-forming cloud population.

Finally, we note that both cloud catalogues rely on CO as a tracer of molecular gas and our mass estimates above rely on assumptions about the CO-to-H$_2$ conversion factor in these galaxies.  Moreover, these two studies use two different line tracers ($J=2-1$ for M33 and $J=1-0$ for M31), so the catalogues inherit potential biases from relying on CO emission as a proxy for star forming gas.  Our study, however, relies primarily on the locations of the CO emission and not on its brightness, so concerns about conversion factor variations \citep[][]{2011Leroy} will manifest as uncertainties in the mass completeness limit. Similarly, systematic variations in the CO line ratios with galaxy and environment \citep[e.g.][]{2020Koda,2021denbrok,leroy22} may represent a caution for comparing the M33 and M31 results.  Our analysis relies on whether a given cloud is detected or not, so such concerns will manifest as changes in the true completeness limits relative to those determined by false source injection of CO sources.  Given the magnitude of these variations observed in the literature, we expect that the typical change in completeness limit would be about 0.2 dex, though these variations are typically measured in more massive systems than M33.  This effect is relatively small compared to the range of cloud masses probed ($\sim 2$~dex) so is not likely to dramatically change the results.

Despite careful measures of completeness for both the clouds and clusters, the lack of constraints on the mass fraction of stars found in bound clusters ($f_\mathrm{clust}$) and the relationship between cloud mass and cluster mass \citep{krumholz19} precludes a clear answer to how tightly the two populations relate to each other.  We proceed assuming the populations are comparable and our later results do not contradict this assumption.  

\section{The Cloud-Cluster Population in M33}\label{m33analy}
Utilizing the high-quality data of clusters (PHATTER) and GMCs (ALMA ACA) in M33, we determine how these clusters and GMCs are correlated. First, we find the separations between the clusters and clouds and determine if they correlate with the cluster's age, as would be expected in the standard model of cloud and cluster evolution. We then refine this measurement with a two-point correlation analysis of the individual objects and a cross-correlation analysis of the clusters and GMCs. This cross-correlation analysis indicates the degree to which clusters are statistically associated with GMCs. We then look at how the properties of the clusters are related to the properties of the associated GMC, with a particular focus on the angle at which the clusters leave their associated clouds. Finally, we determine the lifetime of the GMCs in M33 by using the method of \cite{Kawamura2009}, which is based on the spatial overlap of young clusters.  These results depend on the completeness limits of the contributing catalogues, which we discuss in more detail in Section \ref{sec:completeness}.

\subsection{Cloud-Cluster Spatial Offsets}\label{m33methods1}

To compare the properties of the GMCs and clusters, we must first find their locations in the plane of M33's disk. As seen in Figure \ref{fig:m33dat} the clusters (from PHATTER) and the GMCs (from ALMA ACA) cover different areas. Therefore, we only consider GMCs and clusters in the overlapping survey areas, resulting in a sample of 444 GMCs and 934 of 1214 clusters. Here we treat the SCIMES molecular clouds and PHATTER clusters as point sources. We use the centres of the clusters as the point source location. For the molecular clouds, we use the location of the brightest CO emission (CO peak) as the location of the point sources. Using the orientation parameters from Table \ref{tab:galaxydetails}, we convert celestial coordinates into galactocentric coordinates for each object and measure distances in the plane of the galaxy.

\begin{figure}
	\includegraphics[width=\columnwidth]{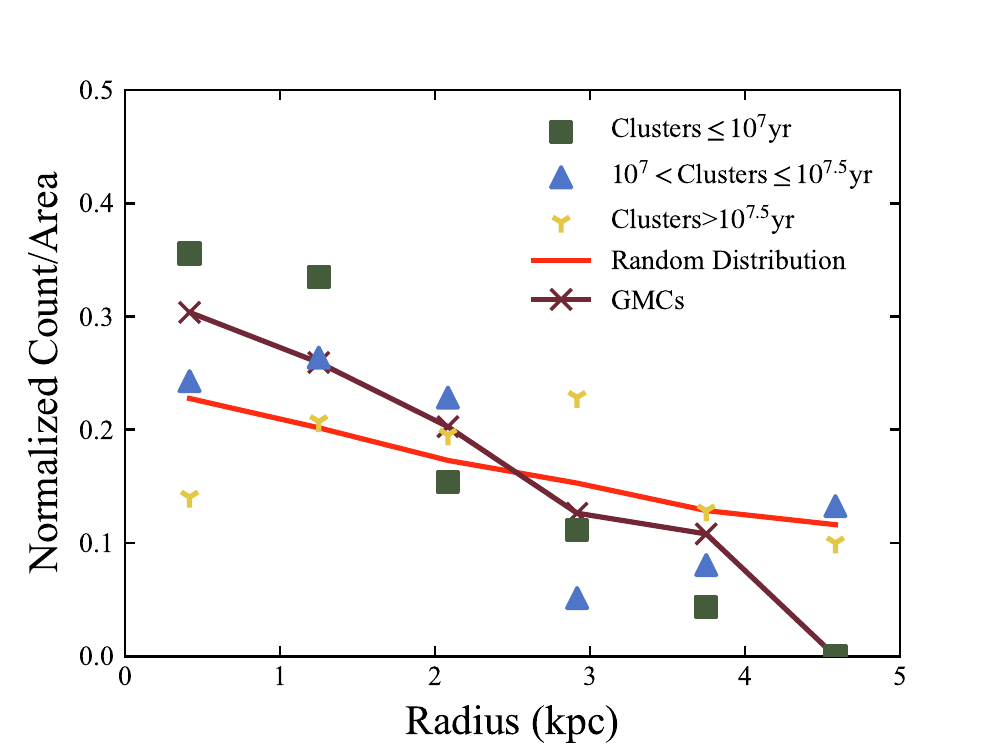}
    \caption{The normalized count per area of the clusters, GMCs, and the generated random clusters. The count is found in galactocentric radial bins and then divided by the area contained in the overlapping survey region in that radial bin. The clusters are split into the three age categories youngest (green squares), medium-aged (blue triangles), and oldest (yellow tri-points). The GMC radial distribution (dark pink line with crosses) is found from the coordinates of the CO peak of each cloud. The random cluster distribution (red line) results from averaging 100 exponential distributions with a scale length fitted from the oldest clusters. The random distribution traces a similar distribution to the oldest clusters in M33.}
    \label{fig:Rdist}
\end{figure}

We then compare the azimuthally averaged radial distributions and generate random distributions that match the radial distributions of the different objects. We use these random distributions to assess the significance of our results. Figure \ref{fig:Rdist} shows the radial distribution for the GMCs and the clusters split into three age categories. We choose three age categories that have distinct relations to the GMCs. The 60 youngest clusters have ages $\leq10^7$~yr, the 93 medium-aged clusters have ages between $10^7$~yr and $10^{7.5}$~yr, and the 781 oldest clusters are $>10^{7.5}$~yr old. 

We assume that GMCs and clusters both follow an exponential distribution for surface density, so that the number in a given radial bin is:
\begin{equation}
    N \sim 2\pi R\, \Delta R\, \exp\left(-\frac{R}{R_d}\right),
\end{equation}
where $R$ is the radial distance from the galactic centre and $\Delta R$ is the width of the radial bin. The different cluster age bins and the GMCs have different scale lengths, $R_d$, that can be seen from the shapes of each radial distribution. We then generate random exponential distributions, selecting the same number of sources (934) in the same overlapping survey area. We generate 100 of these random distributions at each scale length for a range of scale lengths. Then we compared these random distributions to our real distributions and found the real-random distribution pair with the lowest chi-square value. We find a best-fitting scale length of 1.6~kpc for the youngest clusters, 3.2~kpc for the medium-aged clusters, 5.8~kpc for the oldest clusters, and 2.5~kpc for the GMCs. These scale lengths differ from M33's molecular gas scale length of $\approx2.1$~kpc \citep{2014Druard} and stellar scale length of $\approx1.55$~kpc \citep{2009Verley}. The averaged random distribution fitted to the oldest clusters (5.8~kpc) is shown in Figure \ref{fig:Rdist} as a red line. This random distribution based on the oldest clusters is used as our standard reference, but we use the other distributions for our correlation analysis. 

There are many possible explanations for the differing scale lengths. The young clusters have a scale length consistent with the stellar scale length \citep{2009Verley}, and the GMCs have a scale length reasonably close to M33's molecular gas scale length \citep{2014Druard}. However, the medium-aged and oldest clusters have longer scale lengths than the stellar scale length indicating fewer clusters at smaller radii, which is clear from Figure \ref{fig:Rdist}. Therefore, there might be something preventing the older clusters from being identified in the central region of M33. One possible explanation that is discussed in more detail in Section \ref{sec:completeness} is that clusters with young bright stars are easier to visually identify in the crowded central region \citep{2022PHATTERIII}. However, this trend has also been observed in the Milky Way \citep{2018GaiaOC} and in simulations \citep{2008stellarmigration}. Therefore, the more likely explanation is due to clusters being destroyed in the  crowded central region and from clusters migrating to larger radii.

\begin{figure}
	\includegraphics[width=\columnwidth]{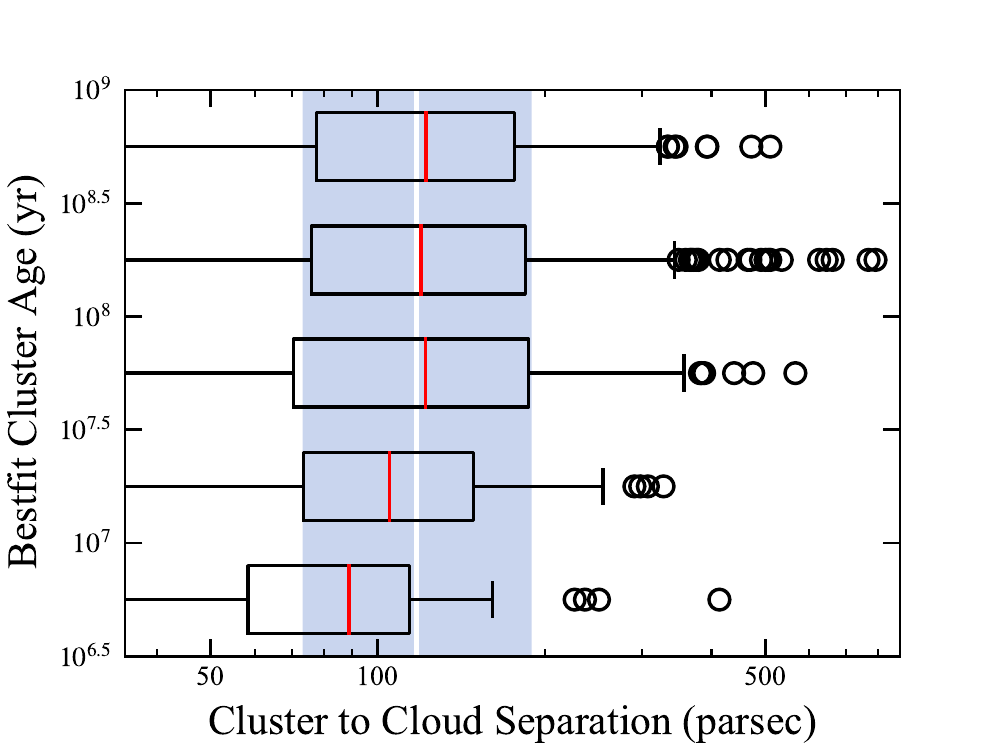}
    \caption{The separation between the clusters and the GMCs based on the age of the clusters in M33. The boxes show the span of the first and third quartiles (interquartile range or IQR) for each age bin that are $10^{0.5}$~yrs wide, with the medians marked with red lines. The error bars extending from the boxes indicate the minimum and maximum values in each bin, excluding the outliers. The outliers, marked with circles, are defined as points outside of $1.5$ times the IQR. The separation between clusters with random positions and the GMCs have their median (white line), and IQR (blue shaded region) plotted. The youngest clusters have a shorter median and a smaller IQR than random clusters. Clusters older than $10^{7.5}$~yrs have medians and IQRs consistent with the random distribution.}
    \label{fig:boxplot}
\end{figure}

We expect that the youngest clusters will be closer to their parent GMCs \citep[e.g.][]{Kawamura2009, 2017CorbM33,2018Grasha,2019Grasha}, which can be shown by comparing cluster age and cluster-cloud separation. We find the closest GMC to each cluster using the galactocentric coordinates. We then measure the physical separation between the centre of each cluster and the nearest molecular cloud CO peak. Figure \ref{fig:boxplot} shows the result of creating a box plot with the separation versus cluster age. Clusters of all ages have typical separations far greater than the 35~pc resolution of the ALMA ACA survey. The youngest clusters have the lowest median separation of 90~pc and the smallest interquartile range (IQR) of 60~pc. With a median separation of 100~pc, the medium-aged clusters are further from GMCs and have an IQR of 80~pc, larger than the youngest clusters. The oldest clusters comprised of three bins in Figure \ref{fig:boxplot} are all quite similar, with medians of 120~pc and IQRs of 100-120~pc consistent with random. The random median (120~pc) and IQR (120~pc) shown in Figure \ref{fig:boxplot} is from the oldest random distribution ($R_d$=5.8~kpc). Using the other random distributions shifts the median and IQR by $\approx10$~pc, which is still most consistent with the oldest clusters. As expected, the clusters start close to a GMC and drift towards randomly distributed as they age. 

\subsection{The Two-Point Correlation Function}\label{2pnt}

\begin{figure}
	\includegraphics[width=\columnwidth]{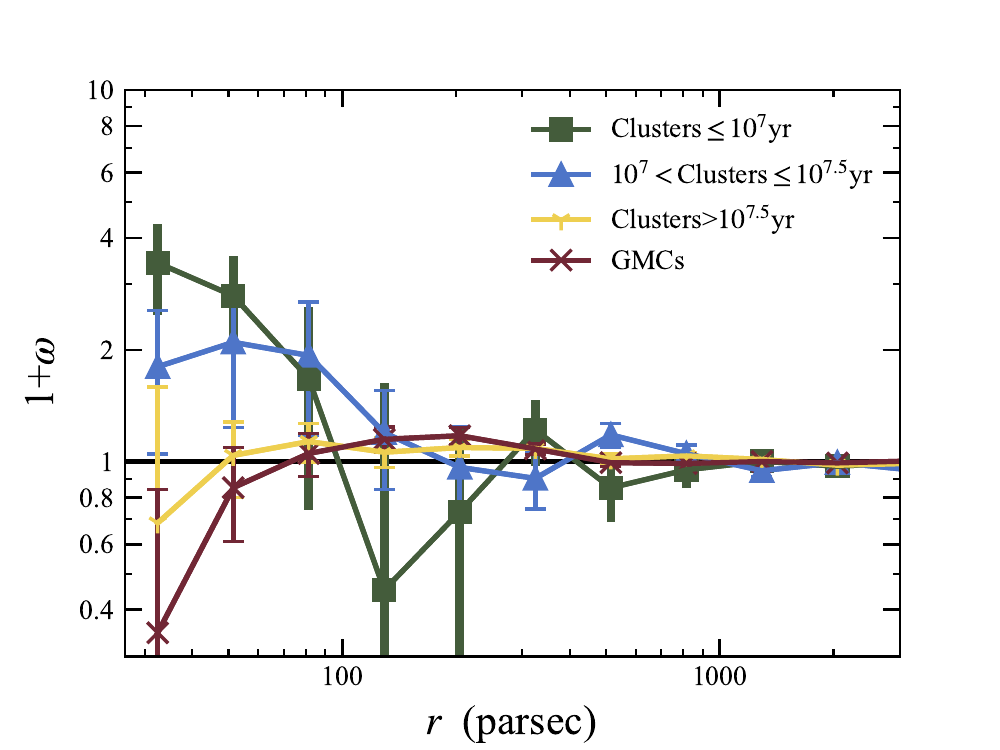}
    \caption{The two-point correlation function, $1+\omega$, at radial separations, $r$ in M33. The two-point correlation is shown for the GMCs as a dark pink line with crosses. The two-point correlation is also shown for the youngest clusters (green line with squares), medium-aged clusters (blue line with triangles), and the oldest clusters (yellow line with tri-points). These two-point correlations are calculated with respect to the 100 random cluster distributions and averaged. The error bars show the standard deviation of 100 two-point correlations. The black line marks uncorrelated. This plot shows that the youngest and medium-aged clusters are correlated, and the GMCs are anti-correlated at small radii. At larger radii, all the groups become uncorrelated.}
    \label{fig:twopnt}
\end{figure}

We now further analyze the spatial properties of our distributions using the two-point correlation function that quantifies the amount of clustering on different spatial scales \citep{1980Peebles}. In general, the two-point correlation function describes the probability of finding an object in two volume elements separated by $r$. The two-point correlation of a real data set can then be compared to the two-point correlation of a random distribution to find the excess probability of spatial correlation \citep{1980Peebles}. We use a slight variation on the two-point correlation function since we are working in the plane of the disk. Therefore, we use $\omega(r)$ to indicate the excess probability that an object will be found at a distance $r$ from another object of the same type compared to a random distribution. $1+\omega=1$ indicates a random uncorrelated distribution, $1+\omega>1$ indicates a correlated distribution, and $1+\omega<1$ indicates an anti-correlated distribution. $\omega(r)$ is calculated in radial bins separated by $\log_{10} (r)=0.2$. For each radial bin, we count the number of real catalogue pairs $DD$, random catalogue pairs $RR$, and pairs of one real and one random object $DR$. Then we use the \cite{1993ApJ...412...64L} estimator in the same form as \cite{2022Turner}:
\begin{equation}
    \omega(r)=\frac{1}{RR}\left[DD\left(\frac{N_R}{N_D}\right)^2-2DR\left(\frac{N_R}{N_D}\right)+RR\right],
\end{equation}
where $N_R$ is the total number of objects in the random catalogue, and $N_D$ is the total number of objects in the real catalogue. This process is repeated for each object type using their respective 100 random exponential distributions. Figure \ref{fig:twopnt} shows the average two-point correlation function for the GMCs and the three cluster age categories. The standard deviation in two-point correlations from the random distributions is shown in Figure \ref{fig:twopnt} as error bars. 

The youngest and medium-aged clusters are correlated at separations smaller than $\sim 100$~pc. The oldest clusters are uncorrelated at all scales. GMCs show anticorrelation on small ($<50$~pc) scales, which we attribute to the object identification algorithm. When the SCIMES algorithm defines the local maxima of GMCs, it requires a minimum spatial separation between the maxima, which is set to 50 pc (Koch et al. in prep). As the radial separation increases beyond 100~pc, all catalogues tend toward $1+\omega=1$, which indicates they are uncorrelated on large scales. The youngest clusters show modest anticorrelation on medium scales (200 pc) and are uncorrelated at larger scales. This anticorrelation is likely the consequence of strong correlation at small scales ($<100$~pc). Not shown in Figure \ref{fig:twopnt} is the two-point correlation of the YSCCs from \citet{2017CorbM33} because we find no significant correlation at any scale. 

We also tried to test the effects of completeness by removing clusters below a certain mass. We removed the clusters below $10^{3.5}$~M$_\odot$ where the cluster catalogue is $\approx$90\% complete. This high level of completion should eliminate the effects of difficulty identifying older clusters and in crowded regions. There are fewer clusters which makes the correlation structure less consistent. However, the trend for decreasing correlation strength with age is still apparent. Therefore, we assume the change in completeness due to crowding and cluster ageing will not significantly impact our results. Performing a similar test for the GMCs, removing the GMCs below $3.6\times10^4$~M$_\odot$ leaving only the GMCs that are $\approx$90\% complete. Removing these lower-mass clouds has very little effect on the two-point correlation of the GMCs. \citet{2018Grasha,2019Grasha} found that removing lower mass clusters and GMCs resulted in higher correlation magnitudes. We see this effect with the clusters but not with the GMCs. 

The results of our two-point correlation analysis are consistent with what other studies have found. \citet{2018Grasha,2019Grasha} and \citet{2022Turner} all found that young clusters (<10~Myr) have stronger correlation than the older clusters. While the main results are consistent, there are two major differences between our results and previous studies. In NGC 7793, \cite{2018Grasha} found the two-point correlation using a catalogue of 293 clusters using the same estimator. However, \citet{2018Grasha} find a larger magnitude of correlation and the clusters remain correlated until separations of approximately 1000~pc instead of 100~pc. \citet{2019Grasha} and \citet{2022Turner} found the same difference in magnitude and scale in the galaxy M51 and 11 PHANGS galaxies, respectively. This difference likely comes from the inherent clustering of stars in a galaxy that we have tried to account for by using exponential random distributions fitted to our clusters. 

Another important result that is consistent between our results and \citet{2018Grasha,2019Grasha} and \citet{2022Turner} is that the GMCs are much closer to a random distribution than the clusters. If each GMC produced only a single cluster it would be expected for the correlation structure of the GMCs and young clusters to match. The strong correlation seen in the youngest clusters and the lack of correlation in GMCs could suggest that GMCs produce multiple clusters. However, we find that none of our GMCs are overlapping with several of the youngest clusters, which does not fully rule out this possibility since clusters drift. Another possible solution is that enough GMCs are quickly destroyed by young clusters that the correlation structure of the GMCs is erased. Finally, a more mundane solution would be to attribute the lack of correlation to the cloud identification algorithm, which suppresses the correlation structure in the molecular clouds at short scales (Figure \ref{fig:twopnt}) by forcing them into discrete, well separated units. The resolution of the ALMA ACA survey could prevent the cloud identification algorithm from distinguishing between a complex of smaller clouds and one large cloud. We will address this possibility by analyzing the correlation of all of the $^{12}$CO emission without cloud decomposition in Peltonen et al. (prep).
 
\subsection{The Cloud-Cluster Cross Correlation}\label{cross}
\begin{figure}
	\includegraphics[width=\columnwidth]{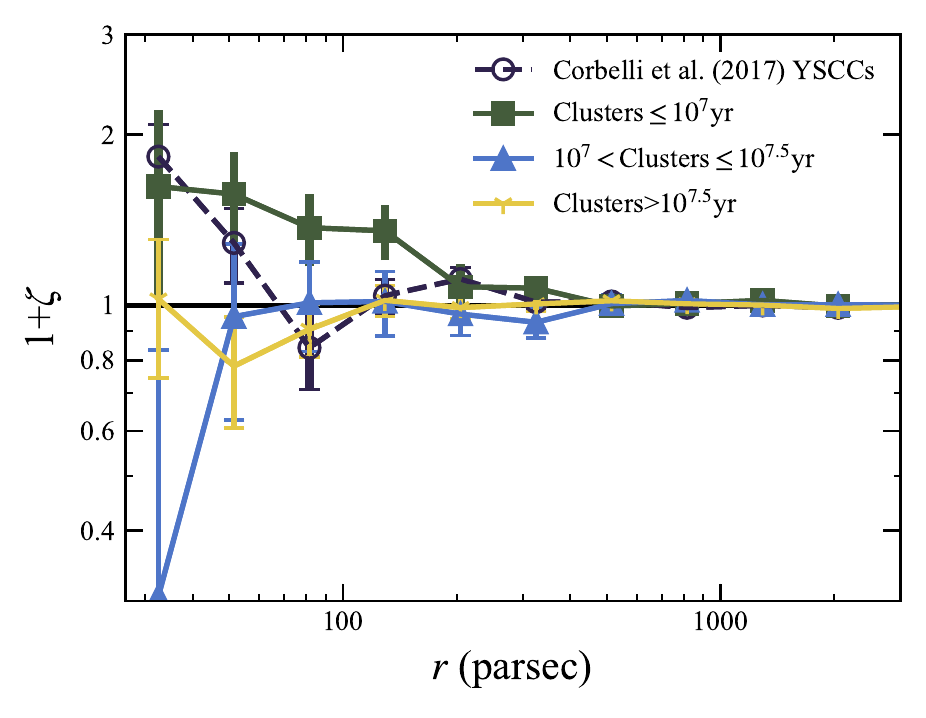}
    \caption{The cross-correlation function, $1+\zeta$, at radial separations, $r$ between the clusters and GMCs in M33. The cross-correlation is shown for GMCs correlated with the youngest clusters (green line with squares), medium-aged clusters (blue line with triangles), and the oldest clusters (yellow line with tri-points). The blue dotted line with circles shows the cross-correlation between the \citet{2017CorbM33} YSCCs and GMCs. These cross-correlations have been found with respect to the 100 random cluster and GMC distributions and averaged. The error bars show the standard deviation of 100 cross-correlations. The black line marks uncorrelated. This plot shows that the youngest clusters are correlated with the GMCs at small radii.}
    \label{fig:crosscor}
\end{figure}

We want to understand how cluster distributions are related to the GMC distribution at different scales, which can be shown using the cross-correlation function. The cross-correlation function, $\zeta(r)$ indicates the excess probability that two data sets are jointly clustered more than two random data sets at a distance $r$. As with the two-point correlation function, $1+\zeta=1$ marks the boundary between correlated and anti-correlated. We use the cross-correlation function to find the correlation between GMCs and clusters in the three age categories. To estimate this cross-correlation, we use the three random cluster distributions and the random GMC distributions. Then by repeating this process with all 100 respective random distributions, we find an average cross-correlation. For each radial bin, we find the number of real cluster GMC pairs $D_{sc}D_{gmc}$, real cluster random GMC pairs $D_{sc}R_{gmc}$, random cluster real GMC pairs $R_{sc}D_{gmc}$, and random cluster random GMC pairs $R_{sc}R_{gmc}$. Again, we use the \cite{1993ApJ...412...64L} estimator in the same form as \cite{2022Turner}:
\begin{align}
\begin{split}
    \zeta=\left(\frac{N_{R_{sc}}N_{R_{gmc}}}{N_{D_{sc}}N_{D_{gmc}}}\frac{D_{sc}D_{gmc}}{R_{sc}R_{gmc}}\right)-\\
    \left(\frac{N_{R_{sc}}}{N_{D_{sc}}}\frac{D_{sc}R_{gmc}}{R_{sc}R_{gmc}}\right)-\left(\frac{N_{R_{gmc}}}{N_{D_{gmc}}}\frac{R_{sc}D_{gmc}}{R_{sc}R_{gmc}}\right)+1,
\end{split}
\end{align}
where $N_{D_{sc}}$ is the number of real clusters, $N_{D_{gmc}}$ is the number of real GMCs, $N_{R_{sc}}$ is the number of random clusters, and $N_{R_{gmc}}$ is the number of random GMCs. Figure \ref{fig:crosscor} shows the result of finding the average cross-correlation between the GMCs and the clusters of the three age categories. The standard deviations of the 100 cross-correlations are shown in Figure \ref{fig:crosscor} as error bars. The youngest clusters are correlated with GMCs at small radial separations. The medium-aged and oldest clusters are uncorrelated at most separations. In the first bin, the medium-aged clusters are anti-correlated. However, this could be explained by the large standard deviations, smaller number of sources at small separations, and extinction from the GMCs. The older clusters would be less subject to the extinction since older clusters have larger scale heights. The youngest clusters are likely also affected by this extinction. The correlation at small scales is expected to be quite strong, which could be partially overcoming the extinction, and the missing embedded clusters would likely amplify the correlation strength at these small separations. Removing the clusters and GMCs below the 90\% completeness limits slightly reduces the correlation magnitudes. However, the overall shape of the correlations remains unchanged and our conclusions are the same. 

The correlation between the \citet{2017CorbM33} YSCCs in our survey area (244 out of 630) and the GMCs is similar to our youngest clusters. We have found this correlation with respect to a random distribution with a scale length of 2.9~kpc, which was found in the same way as the other random exponential distributions. The YSCCs become uncorrelated at a smaller scale than our youngest clusters. The YSCCs also have a greater magnitude of correlation at the smallest scale. This increased correlation could be due to the very young $\leq$3~Myr clusters that cannot be identified in the optical. 

Fewer studies have analyzed cross-correlation than two-point correlation. However, our results appear consistent with what \cite{2022Turner} found for a sample of 11 PHANGS galaxies. There is some variation over the 11 galaxy sample, but \cite{2022Turner} find that clusters younger than 10~Myr have similar correlation magnitudes to what we have found. There is still a difference in random catalogues, but \cite{2022Turner} find that older clusters are typically much less correlated with GMCs, similar to our analysis. This correlation between young clusters and GMCs allows for two possibilities. The first possibility is that clusters are still nearby to their parent GMC, which means that the cluster has not had sufficient time to fully destroy the progenitor cloud. The second possibility is that the cluster has destroyed its true progenitor cloud but, is still associated with a nearby GMC. As discussed in Section \ref{2pnt} our cloud identification algorithm might be identifying complexes of multiple GMCs as a single larger GMC which would erase the two-point correlation of the GMCs. Therefore, it would be difficult to distinguish between these two possibilities using these methods. We note that, unlike the real GMC catalogue, objects in our random GMC catalogue are not required to be separated by 50~pc. Therefore, the random GMC catalogue is more correlated at small scales, which could decrease the cross-correlation magnitudes at these small scales.

 \begin{figure}
	\includegraphics[width=\columnwidth]{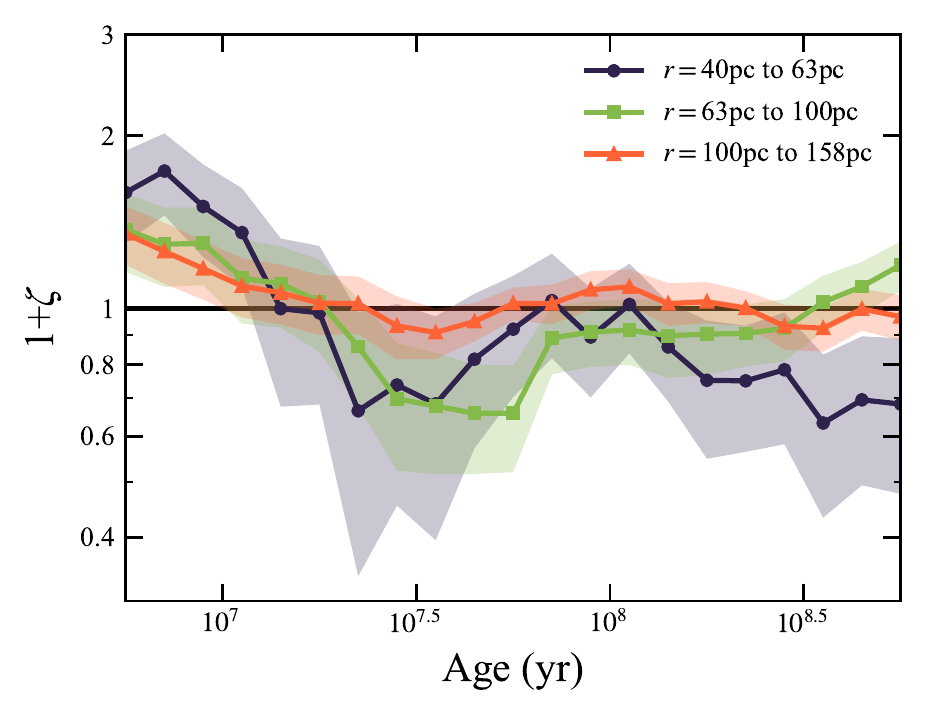}
    \caption{The cross-correlation function, $1+\zeta$, at specific radial bins, versus cluster age in M33. The second (dark blue line with dots), third (light green line with triangles), and fourth (orange line with squares) bins from Figure \ref{fig:crosscor} are plotted. Where the age bins are 0.5 dex wide. The cross-correlations have been found with respect to the 100 random cluster and GMC distributions and averaged. The shaded regions show the standard deviation of the 100 cross-correlations. The black line marks uncorrelated. The cross-correlation between clusters and GMCs decreases with age.}
    \label{fig:crosscoronebin}
\end{figure}

We now determine at what age clusters transition from correlated to uncorrelated with GMCs. Figure \ref{fig:crosscoronebin} shows the cross-correlation at specific radial bins broken into smaller, overlapping age groups. The age groups are logarithmically spaced with a width of 0.5 dex, starting with $10^{6.5}-10^7$~yr and increasing by a factor of 0.1 dex. We choose overlapping age bins to increase the number of sources and reduce noise. The radial bins used (second, third, and fourth bins from Figure \ref{fig:crosscor}) are where the youngest clusters are correlated with GMCs and the medium-aged clusters are not anti-correlated. To find the scale length of the random distribution for each overlapping bin we assume the scale length increases linearly with age. Figure \ref{fig:crosscoronebin} shows that the clusters begin with being correlated with GMCs at these small separations, and then at $\approx 18$~Myr, all bins are consistent with being uncorrelated. The three radial bins in Figure \ref{fig:crosscoronebin} behave similarly before $10^7-10^{7.5}$~yr, but with different amplitudes of correlation. The longest radial bin (100~pc to 158~pc) trends directly to uncorrelated at $\approx 18$~Myr while the shorter bins become anti-correlated. This anti-correlation could be due to the larger correlation at younger ages or from extinction from GMCs.

\subsection{Properties Of Associated Clusters} \label{properties}

Since the molecular cloud population is the site of star formation, the resulting star formation properties and cluster population should depend on the properties of the progenitor clouds \citep{kruijssen12, krumholz19}. We look at the properties of the clusters younger than 10~Myr that are likely still correlated with GMCs. We find, similar to other studies \citep{2017CorbM33,2018Grasha,2019Grasha}, that the properties of the nearest cloud (like mass, radius, and surface density) have no significant correlation with the mass of the clusters produced. This lack of correlation could mean cluster properties are not determined by the properties of the cloud. However, this lack of correlation can also be explained by the evolution of GMCs. It has been suggested that GMCs continually accrete additional gas throughout their lifetimes \citep{2009Fukui,2012Gratier}. Then once the GMC has produced a sufficient number of stars, the GMC is dispersed through stellar feedback. Therefore, the properties of GMCs are not constant, and the properties of the nearest GMC to a cluster are unlikely to be the same as the progenitor cloud's properties at the time of formation.

\begin{figure}
	\includegraphics[width=\columnwidth]{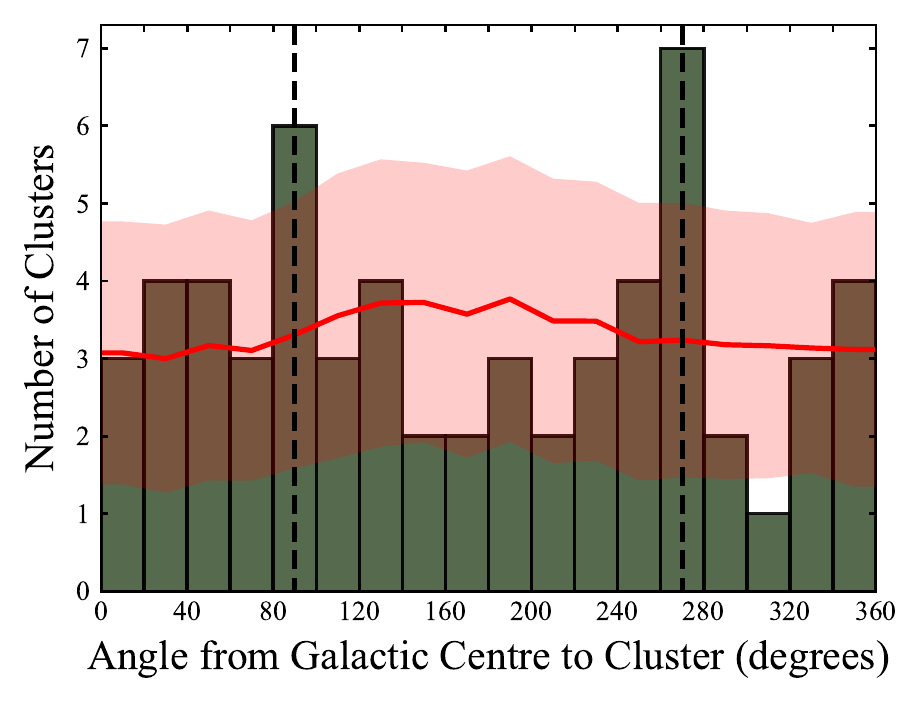}
    \caption{The angular distributions of real and random young clusters ($\leq$10~Myr) in M33. The green histogram shows the distribution of angles between pairs of young clusters and their nearest GMC. The angle is defined from the GMC to the young cluster with 0$^\circ$ pointing towards galactic centre. The mean random angular distribution from the 100 random young cluster positions is shown as a red line. The red shaded region shows the standard deviation in the 100 random angular distributions. The angular bins are 20$^\circ$ wide. The vertical dashed lines indicate 90$^\circ$ and 270$^\circ$. The real angular distribution of the young clusters is slightly different from random clustering.}
    \label{fig:AngleAllClus}
\end{figure}

One property that will be more difficult to erase via feedback is the direction a cluster leaves its progenitor cloud. If a cloud forms multiple clusters or clouds exist in complexes, the direction of the correlated young cluster ($\leq$10~Myr) will be preserved. However, if the progenitor cloud is destroyed and there are no GMCs near the progenitor cloud then the direction will not be preserved. We select the nearest young cluster-cloud pairs and find the angle from the cloud to the cluster. We define the angle to start from 0$^{\circ}$ pointing from the molecular cloud to the galactic centre and with 90$^\circ$ pointing in the direction of galactic rotation. The histogram in Figure \ref{fig:AngleAllClus} shows this cloud to cluster angle broken into 20$^\circ$ bins. We then found the angles from our molecular clouds to the 100 random young cluster distributions ($R_d$=1.6~kpc). The 100 angle distributions are then averaged, which is shown in Figure \ref{fig:AngleAllClus} as a red line. The red shaded region in Figure \ref{fig:AngleAllClus} shows the standard deviation in the 100 random angular distributions. Young clusters are marginally more likely to be at 90$^\circ$ (in the direction of galactic rotation) and 270$^\circ$ (in the opposite direction of rotation). We recognize that the significance of this result is weak, but we find it notable that the peaks in the real angular distributions lie in the 90$^\circ$ and 270$^\circ$ directions, though nothing in the analysis favours these directions. Even after varying the bin positions and widths, the peak at 270$^\circ$ is persistent. However, the strength of the peak at 90$^\circ$ is diminished in some variations of binning. 

\subsection{Molecular Cloud Lifetimes}

Using our cluster ages, we make an estimate of the total molecular cloud lifetime following a similar procedure as used in \citet{Kawamura2009}. If we assume clusters are formed at a constant rate, the fraction of clusters in a phase represents the time spent in that phase. We divide cloud lifetime into two stages: $\tau_\mathrm{dark}$ where a cloud shows no association with any potential disrupting cluster and $\tau_\mathrm{fb}$ where a stellar source could be providing feedback. Then, $\tau_\mathrm{GMC} = \tau_\mathrm{dark} + \tau_\mathrm{fb}$. 

To determine the length of $\tau_\mathrm{fb}$ we only use the young clusters ($\leq$10~Myr) from \citet{2022PHATTERIV} that are believed to be long-lived. We find the fraction of our youngest clusters overlapping with their parent GMC. We use this fraction of association to estimate how long clusters spend with their GMC. This fraction of 10~Myr will be known as $\tau_{fb}$, the feedback phase. We find that $N_\mathrm{overlap} =23$ out of $N_\mathrm{total} = 60$ young clusters are associated with molecular clouds. Based on the fraction of associated young clusters, we have 
\begin{equation}
    \tau_\mathrm{fb} = 10~\mathrm{Myr}\left(\frac{N_\mathrm{overlap}}{N_\mathrm{total}}\right)\approx \mathrm{4~Myr}.
\end{equation}
However, due to visual extinction, we are likely missing some clusters younger than $3$~Myr. If we assume we are missing every deeply embedded cluster $<3$~Myr ($\approx$30\%), then   $\tau_\mathrm{fb}\approx 6$~Myr. Therefore, we estimate the feedback phase to be 4-6~Myr.

We can now estimate the total lifetime of the GMCs by finding the fraction of GMCs in the feedback phase. This fraction of GMCs must include not only the long-lived clusters but also other sites of high-mass star formation that are visible but not classified as clusters in \citet{2022PHATTERIV} (e.g., OB associations). Then we have:
\begin{equation}
\frac{\tau_\mathrm{fb}}{\tau_\mathrm{dark}+ \tau_\mathrm{fb}} = \frac{N_\mathrm{GMC,fb}}{N_\mathrm{GMC}}
\end{equation}
where $N_\mathrm{GMC,fb}$ is the total number of GMCs experiencing feedback. Koch et al. (in prep) identified 217 GMCs associated with recent high-mass star formation through a visual inspection which includes many of the clusters from \citet{2022PHATTERIV}. However, this inspection was done without the \cite{2022PHATTERIII} ages for the long-lived clusters. We visually inspect the 217 GMCs with visible clusters and find that 56 contain only an old cluster (>10$^7$yr). These clusters that are much older than $\tau_\mathrm{fb}$ are likely not in the same plane as the GMC and do not represent a GMC in the feedback phase. The Koch et al. (in prep) GMCs without the old clusters leave $N_\mathrm{GMC,fb} =161$ out of $N_\mathrm{GMC}=444$ GMCs in the feedback phase. Therefore, the feedback phase represents $\approx$35\% of the total lifetime of GMCs in M33, giving a lifetime of $\tau_\mathrm{GMC} = \tau_\mathrm{dark}+ \tau_\mathrm{fb} = N_\mathrm{GMC} \tau_\mathrm{fb}/N_\mathrm{GMC,fb} = $11-15~Myr. However, this lifetime estimate is sensitive to the visual cluster identification and the removal of older clusters. If the old clusters are not removed, the lifetime estimate would be 8-12~Myr. Regardless of this sensitivity, our GMC lifetime estimate is consistent with the short lifetimes found by previous studies \citep[e.g.][]{Kawamura2009, 2007Blitz, 2012Miura,2017CorbM33,2019Kruijssen,2020Chevance,2021Kim,2022Pan, 2022kim}.

\section{Cloud-Cluster Population in M31} \label{m31analy}
We now analyze the clusters (PHAT) and GMCs (CARMA) in M31 and compare these results to those found in M33. However, the different conditions of M31 only allow for certain results to be compared. M31 has a greater inclination (Table \ref{tab:galaxydetails}) than M33, making photometry more complex and limiting the age estimates for clusters. In addition, the limited survey area and the more defined rings of M31 make creating random catalogues more difficult. Despite these differences, we find the separations between the clusters and GMCs in M31 and compare them to M33's separations, which depend on cluster age. We then find the two-point correlation of the individual objects, which can confirm some of the properties of our methods. Finally, we find the cross-correlation between the clusters and GMCs in M31 to confirm that cross-correlation depends on cluster age.

\subsection{Cloud-Cluster Spatial Offsets}
Similar to M33, in section \ref{m33methods1}, we want to find the positions of the clusters and GMCs on the disk where the survey areas overlap. Including objects where PHAT and CARMA data overlap leaves 480 of 1249 clusters and all 251 GMCs. We can then use the orientation parameters of M31 from Table \ref{tab:galaxydetails}, to convert the celestial coordinates to galactocentric coordinates. 

\begin{figure}
	\includegraphics[width=\columnwidth]{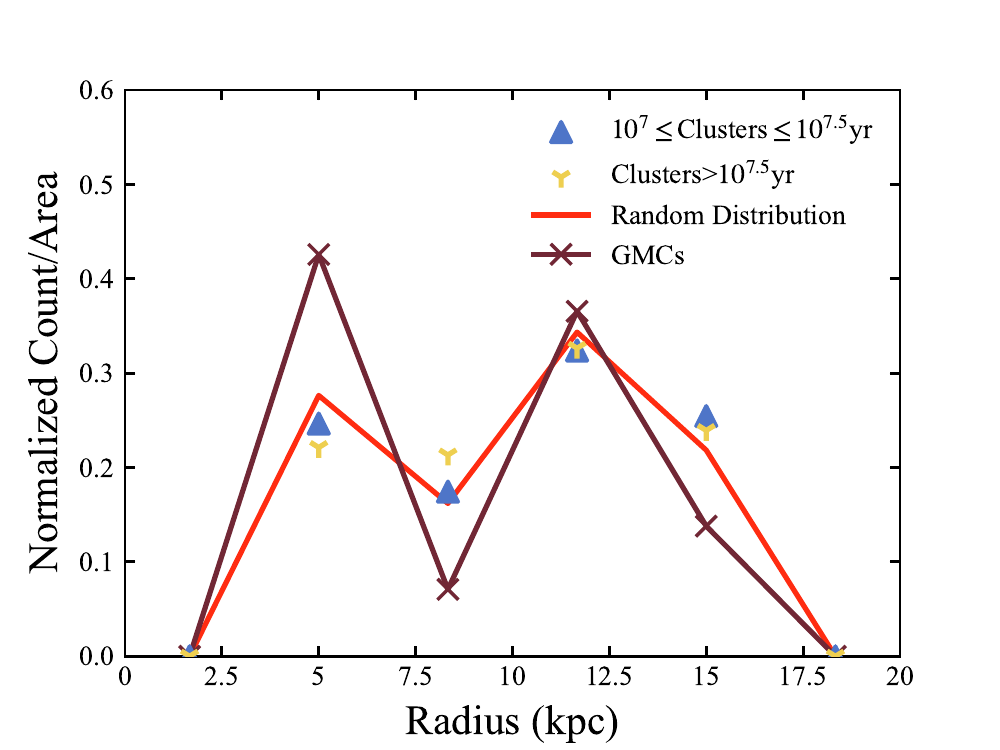}
    \caption{The normalized count per area of the clusters, GMCs, and the generated random clusters. The count is found in galactocentric radial bins and then divided by the area contained in the overlapping survey region in that radial bin. The clusters are split into the two age categories medium-aged (blue triangles) and oldest (yellow tri-points). The GMC radial distribution (dark pink line with crosses) is found from the coordinates of the CO peak of each cloud. The random cluster distribution (red line) results from averaging 100 distributions generated from WISE. The random distribution traces a similar distribution to the clusters in M31.}
    \label{fig:Rdist31}
\end{figure}

The radial distribution of M31 in Figure \ref{fig:Rdist31} looks very different than the radial distributions of M33 (Figure \ref{fig:Rdist}). The GMCs and the clusters split into age categories follow a double peak structure offset from zero. The double peak comes from the odd shape of the CARMA survey area and because the survey area is centred on the star-forming rings of M31. It is also notable that M31 is much larger than M33. Therefore, the clusters and GMCs span a much larger area. The analysis of cluster ages in M31 restricted age values to be between $10^7$~yr and $10^{8.5}$~yr. Therefore, we choose only two age groups the 96 medium-aged clusters have ages $\leq10^{7.5}$~yr and the 384 oldest clusters with ages $>10^{7.5}$~yr. This more limited age range prevents a completely parallel analysis of M33.

Generating random distributions in the same way as M33 by fitting an exponential distribution to the clusters in the survey area yields a flat distribution ($R_d=\infty$). This is likely a consequence of the defined ring structure present in the survey area. A flat distribution does not replicate the observed radial distributions of our clusters. To create a random distribution that matches the defined rings, we must sample a distribution with a similar structure. Therefore, we use the full-galaxy 22$\mu$m image constructed by \citet{2019Leroy} using data from the \textit{Wide-field Infrared Survey Explorer} (WISE) satellite \citep{2010Wright}. We azimuthally average the emission in the WISE filters to create a distribution that can be sampled. The filter W4 (22~$\mu$m) produces random distributions that best match the radial distributions of the clusters and GMCs in M31. The mean of 100 distributions produced from W4 is shown in Figure \ref{fig:Rdist31} as a red line.

\begin{figure}
	\includegraphics[width=\columnwidth]{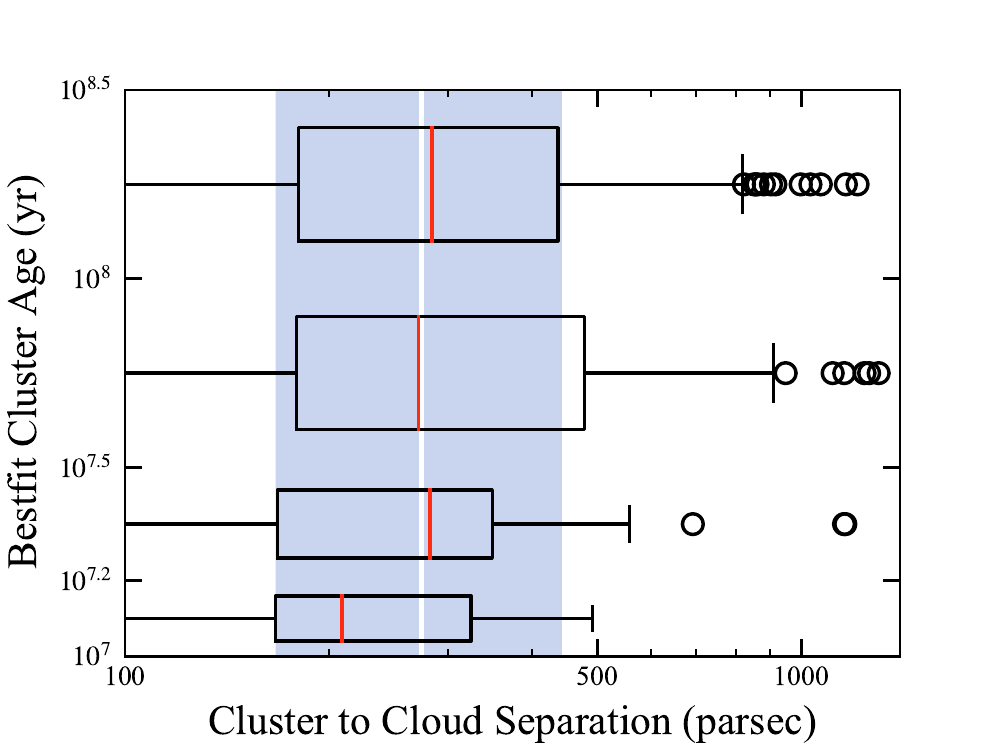}
    \caption{The separation between the clusters and the GMCs based on the age of the clusters in M31. The boxes show the IQR for each age bin, with the medians marked with red lines. The error bars extending from the boxes indicate each bin's minimum and maximum values, excluding the outliers. The outliers, marked with circles, are defined as points outside $1.5$ times IQR. The age bins correspond to the youngest clusters ($10^{7}$~yr-$10^{7.2}$~yr), medium-aged clusters ($10^{7.3}$~yr-$10^{7.5}$~yr), and two bins for the oldest clusters both $10^{0.5}$~yr wide. The separation between clusters with random positions and the GMCs have their median (white line) and IQR (blue shaded region) plotted. The youngest clusters have a shorter median and a smaller IQR than random clusters. Clusters older than $10^{7.5}$~yrs have medians and IQRs consistent with the random distribution.}
    \label{fig:boxplot31}
\end{figure}

Now that we have the positions of the clusters and GMCs in M31, we find the typical separations between clusters and GMCs. Figure \ref{fig:boxplot31} shows the separations between the centre of clusters split by age and the CO peaks of the nearest GMC presented in the same way as Figure \ref{fig:boxplot}. In this section, we split the medium-aged clusters into two smaller age groups since this analysis is less sensitive to the number of objects than the two-point correlation and cross-correlation. The clusters $\leq10^{7.2}$~yr have the lowest median separation of 210~pc and the smallest IQR of 160~pc. The clusters $>10^{7.2}$~yr and $\leq10^{7.5}$~yr have a median separation of 280~pc closer to random (270~pc) but with a smaller IQR of 180~pc than random (280~pc). The oldest clusters have median separations of 270-290~pc and IQRs of 260-300~pc, similar to random. Even with a more limited cluster age range, the same trends are visible in M31 that are seen in M33. We find that the younger clusters are closer to GMCs and the oldest clusters are similar to a random distribution. However, the scale of the separations is larger in M31 because there are fewer objects in a larger area. In M31, the clusters have a density of $\approx 3$~kpc$^{-2}$ and the clouds have a density of $\approx 1$~kpc$^{-2}$. While in M33, the density of clusters is $\approx 36$~kpc$^{-2}$ and the density of clouds is $\approx 17$~kpc$^{-2}$.

\subsection{Two-Point Correlation and Cross-Correlation in M31}
\begin{figure}
	\includegraphics[width=\columnwidth]{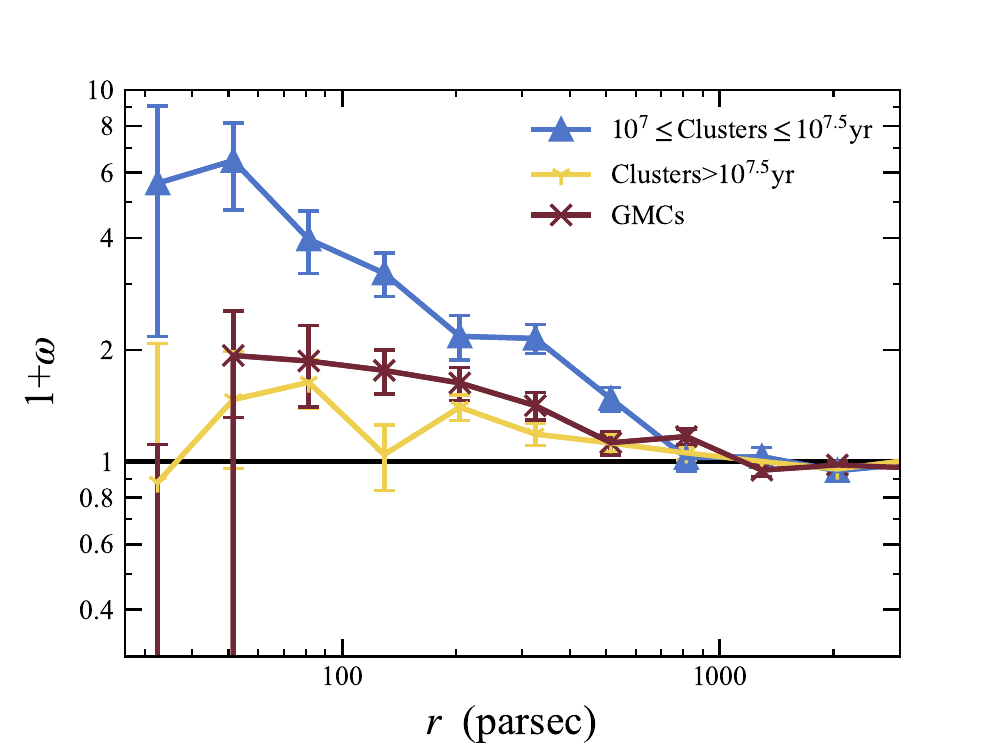}
    \caption{The two-point correlation function, $1+\omega$, at radial separations, $r$ for M31. The two-point correlation is shown for the GMCs as a dark pink line with crosses. The clusters have their two-point correlation shown for the medium-aged clusters (blue line with triangles) and the oldest clusters (yellow line with tri-points). These two-point correlations have been found with respect to the 100 random cluster distributions and averaged. The error bars show the standard deviation of 100 two-point correlations. The black line marks uncorrelated. This plot shows that medium-aged clusters are correlated at small radii. At larger radii, all the groups become uncorrelated.}
    \label{fig:twopnt31}
\end{figure}

We calculate the two-point correlation of the clusters and GMCs in M31. The two-point correlation is found in the same way described in section \ref{2pnt}, showing the excess probability of clustering compared to 100 random distributions. The two-point correlation in different radial bins is shown in Figure \ref{fig:twopnt31} for the GMCs and the clusters in their age categories. The GMCs are anti-correlated at small scales ($<50$~pc) due to the decomposition algorithm. The GMCs are then correlated until $\approx1000$~pc. The oldest clusters have a small amount of correlation until they become uncorrelated at $\approx1000$~pc. The medium-aged clusters have large correlation magnitudes at small separations that decrease until becoming uncorrelated at $\approx1000$~pc. Similar to what we found in M33 we see that the younger clusters have greater correlation magnitudes than the older clusters. The two-point correlation in M31 differs from M33 because the clusters become uncorrelated at much larger scales ($\approx1000$~pc compared to $\approx100$~pc in M33). The two-point correlation in M31 looks more similar to NGC 7793 \citep{2018Grasha}, where objects become uncorrelated at 1000~pc, and the GMCs have small magnitudes of correlation. These two differences could indicate that our random distribution generated from WISE has failed to account for additional large-scale correlation present in M31's defined ring structure. Therefore, WISE W4 may not be the best representative of the full cluster population.

\begin{figure}
	\includegraphics[width=\columnwidth]{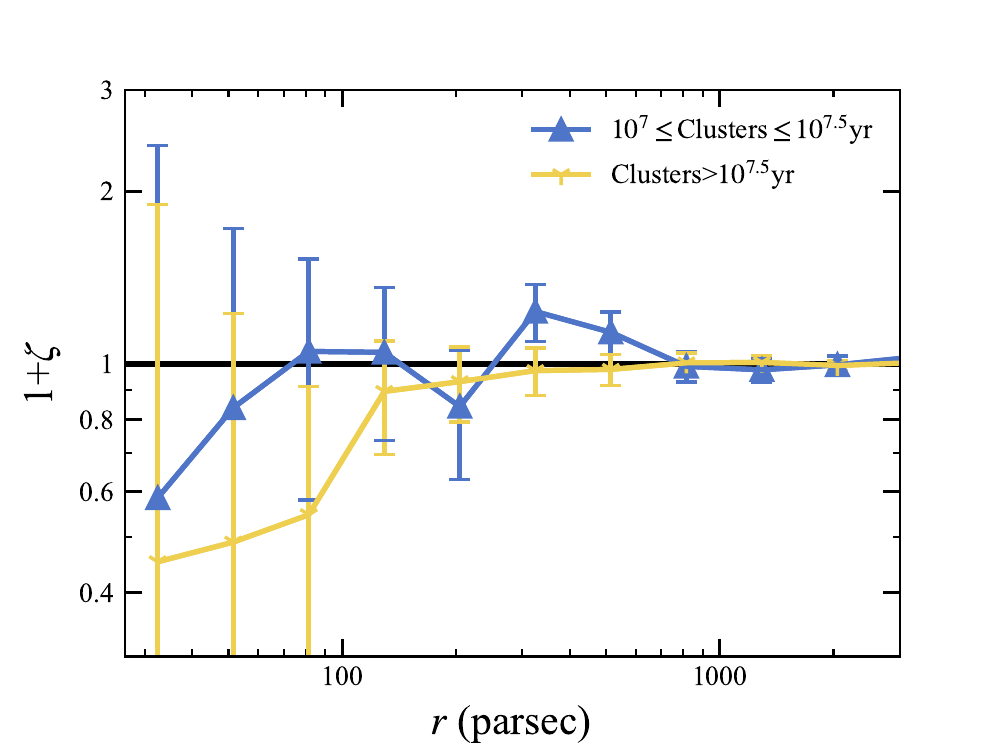}
    \caption{The cross-correlation function, $1+\zeta$, at radial separations, $r$ between the clusters and GMCs in M31. The cross-correlation is shown for GMCs correlated with the medium-aged clusters (blue line with triangles) and the oldest clusters (yellow line with tri-points). These cross-correlations have been found with respect to 100 random cluster and GMC distributions that are both from WISE W4. The error bars show the standard deviation of 100 cross-correlations. The black line marks uncorrelated. This plot shows that there is no significant correlation for clusters of any age in M31.}
    \label{fig:crosscor31}
\end{figure}

As we did for M33, we find the cross-correlation between clusters and GMCs in M31. The methods for finding cross-correlation are the same as presented in section \ref{cross}, finding the excess probability that the clusters are associated with GMCs compared to 100 random distributions. The results of the cross-correlation analysis for M31 are shown in Figure \ref{fig:crosscor31} for the medium-aged clusters and the oldest clusters. There is no significant correlation for clusters at any age, which could be explained by the lack of clusters <$10^7$~yr. Similar to what we find in M33 and consistent with \cite{2022Turner}, there is no correlation for clusters >$10^7$~yr.

\section{Discussion}\label{Discussion}

\subsection{Cluster Drift Speed Estimates}
Our timescale measurements support the idea that the star-formation process occurs rapidly. Once a cluster is formed, it will spend $\approx$4-6~Myr emitting radiation into its parent GMC. Our results also show that GMCs will only survive approximately 11-15~Myr. Therefore, a GMC will experience feedback from clusters for a significant fraction of its lifetime. Our correlation analysis shows that young clusters are correlated with GMCs for $\approx18$~Myr. This loss of correlation likely comes from the original cloud being dispersed and the clusters drifting.

Based on the timescales from our correlation analysis, we can estimate the drift speed, which represents the speed at which the young cluster population decouples from its birth molecular gas. From Figure \ref{fig:crosscor}, we see that the correlation of the youngest clusters and GMCs is no longer present at a scale of 200~pc. Then, from Figure \ref{fig:crosscoronebin}, the correlation between clusters and GMCs disappears at an age of $\approx 18$~Myr. Simply dividing this spatial scale by the temporal scale gives a velocity of $\approx$10~km~s$^{-1}$. It's important to note that this estimate assumes that the lack of correlation at 200~pc is purely due to cluster drift. In reality, this lack of correlation is likely due to a combination of cluster drift and cloud destruction. If a cluster destroys its progenitor cloud, it may still be associated with a more distant GMC in the same complex, which would increase the spatial correlation scale. Therefore, this estimate of drift velocity should be seen as an upper limit. 

\begin{figure}
	\includegraphics[width=\columnwidth]{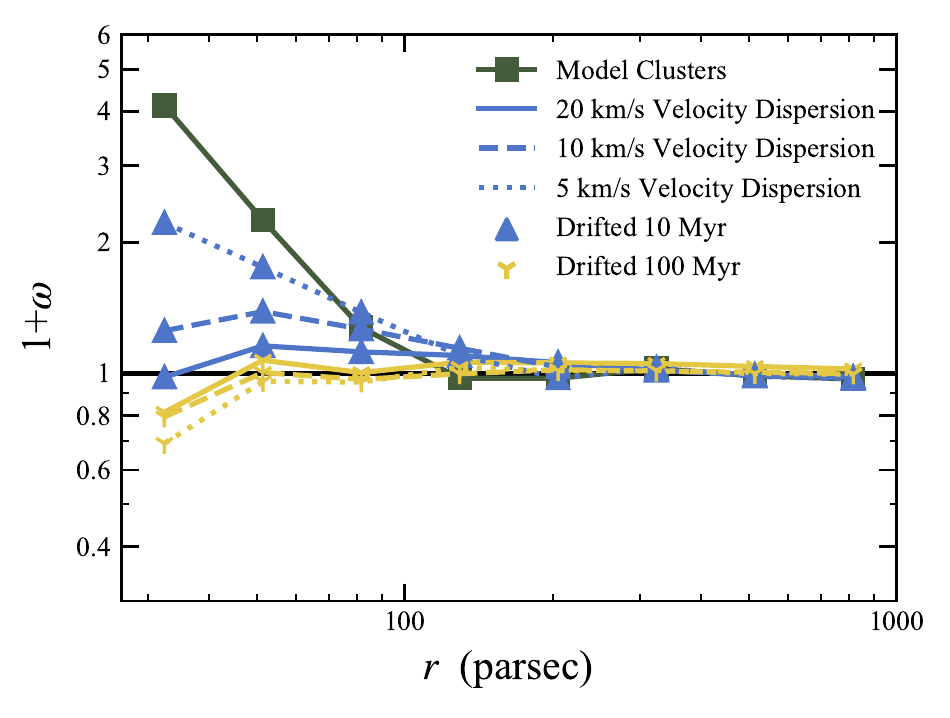}
    \caption{Model two-point correlation function, $1+\omega$, at radial separations, $r$. The two-point correlation of the model clusters to match the youngest clusters in Figure \ref{fig:twopnt} is shown as a green line with squares. The different lines show the average two-point correlation of the different velocity dispersions given to the model clusters. The solid lines show a velocity dispersion of 20~km~s$^{-1}$, the dashed lines show a velocity dispersion of 10~km~s$^{-1}$, and the dotted lines show a velocity dispersion of 5~km~s$^{-1}$. The different colours and symbols show the clusters after 10~Myr (blue triangles) and 100~Myr (yellow tri-points) with their respective velocity dispersion. The black line marks uncorrelated. Our real two-point correlation in M33 falls between the model with a 5~km~s$^{-1}$ and a velocity 10~km~s$^{-1}$ dispersion.}
    \label{fig:gencor}
\end{figure}

We can refine our simple drift model by trying to match the two-point correlation structure seen in M33. We start with a desired correlation model matching the two-point correlation of the young clusters in Figure \ref{fig:twopnt}. A power spectrum can then be generated from Fourier transformation of the model correlation using the relation
\begin{equation}
    P(k) \propto \int_{\mathbb{R}^2} \omega(\mathbf{r}) e^{i\mathbf{k} \cdot \mathbf{r}} d\mathbf{r},
\end{equation}
where $k$ is the wave vector, $r$ is the correlation scale, and $\omega(r)$ is the model correlation at that scale. We then calculate the Fourier transform of the density field as
\begin{equation}
    A(\mathbf{k}) \propto \sqrt{P(k)} e^{i\phi(\mathbf{k})}
\end{equation}
where $\phi(\mathbf{k})$ is a uniformly distributed random phase factor. The density field is then generated and normalized so that, when sampled, it produces a cluster population with a similar correlation structure to the observations. These clusters are then given drift velocities drawn from three different velocity dispersions. These three velocity dispersions (measured in the 2D plane of the galaxy) are 20~km~s$^{-1}$, 10~km~s$^{-1}$, and 5~km~s$^{-1}$. There is about 10~Myr between our youngest and medium-aged clusters and about 100~Myr between our youngest and oldest clusters. Therefore, we let the clusters drift for 10~Myr and 100~Myr and found the two-point correlation for each time and velocity dispersion. This process is repeated 100 times for each velocity dispersion. Figure \ref{fig:gencor} shows the average two-point correlation for each time and velocity dispersion. The 10~Myr drift should have a similar two-point correlation to the medium-aged clusters from Figure \ref{fig:twopnt}, and the 100~Myr drift should be similar to the oldest clusters. All three velocity dispersions seem to replicate the correlation structure of the oldest cluster. However, the magnitude of the correlation of the medium-aged clusters in M33 falls between the 5~km~s$^{-1}$ and 10~km~s$^{-1}$ models. Therefore, this simple model predicts a dispersion velocity of 5-10~km~s$^{-1}$ in M33, which is comparable to the upper limit from the simple model inferred from the age argument above.  

In order to further test the effects of completeness, we assigned masses to our model clusters. We then removed clusters after 10~Myr and 100~Myr of drifting to simulate decreasing completeness for older clusters. The clusters were removed based on the completeness fits from \cite{2022PHATTERIV}. Including this simulated completeness has no noticeable effect on correlation magnitudes. Therefore, we conclude that it is cluster drift and not completeness that results in decreased correlation with cluster age. 

\cite{2018GaiaOC} measured the velocities of open clusters in the Milky Way and found that young clusters (log(age)<7.8) have a velocity dispersion of 10.6~km~s$^{-1}$ among the clusters. This velocity dispersion in the Milky Way increases with age, so this dispersion is consistent with our drift velocity estimation. In M33, \citet{2002Chandar} find that the velocity dispersion of clusters increases with cluster age and for a small sample of clusters <10$^{8.1}$yr find a velocity dispersion of 17~km~s$^{-1}$. Therefore, for our larger sample of younger clusters, a lower velocity dispersion than 17~km~s$^{-1}$ is expected.

Interpreting the physical meaning of this drift velocity can be complicated. Part of this motion represents the random motions of clusters as they are born in a turbulent medium. If the cluster velocity comes from the turbulent motions of the gas, then the dispersion velocity of the gas should be similar to the drift velocity. Koch et al. (in prep) find that the typical line of sight velocity dispersion of CO in molecular clouds to be $\sigma_{v,\mathrm{1D}}\sim 3~\mathrm{km~s^{-1}}$ \citep[see also][]{2003Rosolowsky, 2014Gratier}, which in 2D assuming isotropy would equate to $\sigma_{v,\mathrm{2D}} \sim 2\sigma_{v,\mathrm{1D}} \sim 4.5~\mathrm{km~s^{-1}}$, lower than but of a consistent scale than the drift speed inferred from correlation structure.

\section{Conclusion}\label{Conclusion}
In this paper, we analyze the correlation structure between clusters and GMCs in M33 and M31. We use the PHATTER cluster catalogue from \cite{2022PHATTERIV} with measured CMD ages from \cite{2022PHATTERIII} for M33. In M31, we have the PHAT cluster catalogue from \cite{2015ApJ...802..127J} with measured CMD ages from \cite{2016ApJ...827...33J}. We use the SCIMES algorithm to find GMCs in the 35~pc resolution ALMA ACA $^{12}$CO survey (Koch et al. in prep) in M33 and the 20~pc resolution CARMA $^{12}$CO survey in M31 \citep{calduprimo16,leroy16,schruba19}. We then generate random catalogues of clusters and GMCs that match the radial distributions of the real clusters and GMCs. Using a random catalogue that matches the radial distribution of clusters and clouds removes the spurious signal from large-scale correlations in the cloud-cluster analyses.

In M33 and M31, we find that younger clusters ($\leq$10~Myr) are found at a shorter distance from the nearest GMCs when compared to older clusters (Figures \ref{fig:boxplot} and \ref{fig:boxplot31}). We also find that clusters older than $\approx 30$~Myr have separations from GMCs that matches a random distribution similar to what was found by \citet{2018Grasha,2019Grasha} and \citet{2022Turner}. As seen in Figure \ref{fig:AngleAllClus}, we find that clusters are marginally more likely to drift in the direction of galactic rotation and in the opposite direction of galactic rotation.

The two-point correlation analysis for M33 shows that younger clusters ($<10^7$ yr) are correlated with each other at small scales ($<200$ pc) but older clusters are not (Figure \ref{fig:twopnt}). The same correlation analysis in M31 shows similar but weaker trends due to the more limited cluster age information in M31 (Figure \ref{fig:twopnt31}). The younger clusters likely show a stronger correlation because they have not had sufficient time to drift from other young clusters formed in nearby GMCs. Once the clusters are $\approx$18~Myr old, they no longer show correlation with GMCs. Because of the limited cluster ages, we find no cross-correlation between clusters and GMCs in M31. By comparing the 18~Myr temporal scale to our 200~pc spatial scale, we estimate the drift velocity of the clusters to be 10~km~s$^{-1}$. This is consistent with the drift velocity of 5-10~km~s$^{-1}$ found from comparing our two-point correlation structure to a simple drift model applied to mock data. 

We estimate the time the clusters spend overlapping with GMCs (feedback phase) to be 4-6~Myr in M33. This estimate comes from finding the fraction of the youngest clusters ($\leq$10~Myr) that overlap with GMCs. By finding the fraction of GMCs in the feedback phase, we estimate the total lifetime of GMCs to be 11-15~Myr in M33. 

\section*{ACKNOWLEDGEMENTS}
We thank the anonymous referees for providing comments that improved the quality of this manuscript.
Support for this work was provided by NASA through grant No. GO-14610 from the Space Telescope Science Institute, which is operated by AURA, Inc., under NASA contract NAS 5-26555. JP and ER acknowledge the support of the Natural Sciences and Engineering Research Council Canada (NSERC), funding reference number RGPIN-2022-03499.
EWK acknowledges support from the Smithsonian Institution as a Submillimeter Array (SMA) Fellow and the Natural Sciences and Engineering Research Council of Canada.
L.C.J. acknowledges support through a CIERA Postdoctoral Fellowship at Northwestern University.
AKL gratefully acknowledges support by grants 1653300 and 2205628 from the National Science Foundation and by a Humboldt Research Award from the Alexander von Humboldt Foundation.
ML is supported by an NSF Astronomy and Astrophysics Postdoctoral Fellowship under award AST-2102721.
AS is supported by NASA through grant \#GO-14610 from the Space Telescope Science Institute.

This research made use of NASA’s Astrophysics Data System (ADS) bibliographic services.

We are grateful for the contributions from the open source software community.  This paper made extensive use of software in the \textsc{Astropy} \citep{2013A&A...558A..33A}, \textsc{numpy} \citep{2011CSE....13b..22V}, and \textsc{matplotlib} packages \citep{2007CSE.....9...90H}.

\section*{Data Availability}

This paper makes use of the following ALMA data: ADS/JAO.ALMA\#2017.1.00901.S and 2019.1.01182.S. 
ALMA is a partnership of ESO (representing its member states), NSF (USA) and NINS (Japan), together with NRC (Canada), MOST and ASIAA (Taiwan), and KASI (Republic of Korea), in cooperation with the Republic of Chile. The Joint ALMA Observatory is operated by ESO, AUI/NRAO and NAOJ. The National Radio Astronomy Observatory is a facility of the National Science Foundation operated under cooperative agreement by Associated Universities, Inc. These data are available through the ALMA archive.

Some of the data presented in this paper were obtained from the Mikulski Archive for Space Telescopes (MAST) \url{http://dx.doi.org/10.17909/t9-ksyp-na40}. All other analysis data are available from the authors upon receiving a reasonable request.


\bibliographystyle{mnras}
\bibliography{example}



\appendix


\bsp	
\label{lastpage}
\end{document}